\documentstyle{mn2e}
\input{psfig}

\begin{document}

\def\ergsec{\hbox{erg s$^{-1}$}}
\def\ergcmsec{\hbox{erg cm$^{-2}$ s$^{-1}$}}
\def\countsec{\hbox{counts s$^{-1}$}}
\def\photcmsec{\hbox{photon cm${-2}$ s$^{-1}$}}
\def\degmark{^\circ}
\def \rsun {\ifmmode$R$_{\odot}\else R$_{\odot}$\fi}
\def \nh {N$_{\rm H}$}
\def\ax{$\alpha_{\rm x}$}
\def \hcm {\hbox {\ifmmode $ H atoms cm$^{-2}\else H atoms cm$^{-2}$\fi}}
\def\kms{${\rm km~s^{-1}}$}
\def\PL{${\rm P_{1.4GHz}}$}      
\def\Lel{${\rm L_{e.l.}}$}
\newcommand {\rosat} {{\it ROSAT}}
\newcommand {\einstein} {{\it Einstein }}
\newcommand {\exosat} {{EXOSAT }}
\newcommand {\asca} {{\it ASCA }}
\newcommand {\sax} {{\it BeppoSAX }}
\newcommand {\ginga} {{\it GINGA }}
\newcommand {\ie} {{\it i.e.}}
\newcommand {\cf} {{\it cf }}
\newcommand {\eg} {{e.g.}}
\newcommand {\etal} {et~al. }
\newcommand {\Msun} {{ M$_{\odot}$}}
\newcommand {\degree} {$^{\circ}$}
\newcommand {\sqcm} {cm$^{2}$}
\newcommand {\cbcm} {cm$^{3}$}
\newcommand {\persqcm} {cm$^{-2}$}
\newcommand {\percbcm} {cm$^{-3}$}
\newcommand {\s} {s$^{-1}$}
\newcommand {\gps} {g s$^{-1}$}
\newcommand {\kmps} {km s$^{-1}$}
\newcommand {\lts} {{\it lt}-s}
\newcommand {\yr} {yr$^{-1}$}
\newcommand {\cps} {counts~s$^{-1}$}
\newcommand {\ergs} {erg~s$^{-1}$}
\newcommand {\ergcms} {erg cm$^{-2}$ s$^{-1}~$}
\newcommand {\chisq} {$\chi ^{2}$}
\newcommand {\rchisq} {$\chi_{\nu} ^{2}$}
\newcommand {\cpskeV} {counts s$ ^{-1}$ keV$ ^{-1}$ }
\newcommand {\magg}{\hphantom{$>$}}  
\newcommand {\dig}{\hphantom{0}}
\newcommand {\DXDYCZ}[3]{\left( \frac{ \partial #1 }{ \partial #2 }
                        \right)_{#3}}
\title[\sax observations of 1-Jy BL Lacertae objects - II]
{\sax observations of 1-Jy BL Lacertae objects - II} 

\author[P. Padovani et al.]{Paolo Padovani$^{1,2}$\thanks{Email: padovani@stsci.edu}, Luigi Costamante$^{3,4}$, Paolo Giommi$^5$, Gabriele Ghisellini$^6$, 
\newauthor Annalisa Celotti$^{7}$, Anna Wolter$^{8}$
\\
$^1$ Space Telescope Science Institute, 3700 San Martin Drive, Baltimore MD. 21218, USA \\ 
$^2$ Affiliated with the Space Telescope Division of the European Space 
Agency, ESTEC, Noordwijk, The Netherlands\\
$^3$ Universit\`a degli Studi di Milano, Via Celoria 16, I-20133 Milano, Italy\\
$^4$ Max-Planck Institute f\"ur Kernphysik, Postfach 10 39
80, D-69029 Heidelberg (current address) \\
$^5$ ASI Science Data Center, ASDC, c/o ESRIN, Via G. Galilei, I-00044 Frascati, Italy \\
$^6$ Osservatorio Astronomico di Brera, Via Bianchi 46, I-23807 Merate, Italy\\
$^7$ SISSA, via Beirut 2-4, I-34014 Trieste, Italy\\
$^8$ Osservatorio Astronomico di Brera, Via Brera 28, I-20121 Milano, Italy\\
}

\date{Accepted~~, Received~~}

\maketitle

\begin{abstract}
We present new \sax LECS and MECS observations, covering the energy range $0.1
- 10$ keV (observer's frame), of four BL Lacertae objects selected from the 1
Jy sample. All sources display a flat ($\alpha_{\rm x} \sim 0.7$) X-ray
spectrum, which we interpret as inverse Compton emission. One object shows
evidence for a low-energy steepening ($\Delta \alpha_{\rm x} \sim 0.9$) which
is likely due to the synchrotron component merging into the inverse Compton
one around $\sim 2$ keV. A variable synchrotron tail would
explain why the \rosat~spectra of our sources are typically steeper than
the \sax ones ($\Delta \alpha_{\rm x} \sim 0.7$). The broad-band spectral
energy distributions fully confirm this picture and model fits using a
synchrotron inverse Compton model allow us to derive the physical parameters
(intrinsic power, magnetic field, etc.) of our sources. By combining the
results of this paper with those previously obtained on other sources we
present a detailed study of the \sax properties of a well-defined sub-sample
of 14 X-ray bright ($f_{\rm x}(0.1 - 10$ keV) $> 3 \times 10^{-12}$ erg
cm$^{-2}$ s$^{-1}$) 1-Jy BL Lacs. We find a very tight proportionality between
nearly simultaneous radio and X-ray powers for the 1-Jy sources in which the
X-ray band is dominated by inverse Compton emission, which points to a strong
link between X-ray and radio emission components in these objects.
\end{abstract}

\begin{keywords} galaxies: active -- X-ray: observations

\end{keywords}

\section{Introduction}

BL Lacertae objects constitute one of the most extreme classes of active
galactic nuclei (AGN), distinguished by their high luminosity, rapid
variability, high ($\ga 3$ per cent) optical and radio polarization, radio 
core-dominance,
apparent superluminal speeds, and almost complete lack of emission lines
(e.g., Kollgaard 1994; Urry \& Padovani 1995). The broad-band emission in
these objects, which extends from the radio to the gamma-ray band, appears
to be dominated by non-thermal processes from the heart of the AGN, undiluted
by the thermal emission present in other AGN. Therefore, BL Lacs represent the
ideal class to study to further our understanding of non-thermal emission
in AGN. 

Synchrotron emission combined with inverse Compton scattering is generally
thought to be the mechanism responsible for the production of radiation over
such a wide energy range (e.g., Ghisellini et al. 1998). The synchrotron peak
frequency, $\nu_{\rm peak}$, ranges across several orders of magnitude, going
from the far-infrared to the hard X-ray band. Sources at the extremes of this
wide distribution are referred to as low-energy peaked (LBL) and high-energy
peaked (HBL) BL Lacs, respectively (Giommi \& Padovani 1994; Padovani \& Giommi
1995). Radio-selected samples include mostly objects of the LBL type, while
X-ray selected samples are mostly made up of HBL. 

This scenario makes clear and strong predictions on the X-ray spectra for the
two classes. In the relatively narrow \rosat~band differences between the two
classes became apparent only when very large BL Lac samples ($\sim 50$ per
cent of the then known objects) were considered (Padovani \& Giommi 1996;
Lamer, Brunner \& Staubert 1996). The \sax satellite 
with its broad-band X-ray ($0.1-200$ keV) spectral capabilities, is
particularly well suited for a detailed analysis of the individual X-ray
spectra of these sources.

In Padovani et al. (2001; Paper I) we presented \sax observations of 
seven 1-Jy BL Lacs (plus an additional ``intermediate'' source), six of them
LBL. We discuss here \sax observations of four additional 1-Jy BL Lacs, all
of the LBL type. 

In Section 2 we present our sample, Section 3 discusses the observations and
the data analysis, while Section 4 describes the results of our spectral fits
to the \sax data. Section 5 deals with the \rosat~data for our sources, while
Section 6 presents the spectral energy distributions. Section 7 discusses the
X-ray properties of all 1-Jy BL Lacs observed by \sax, while Section 8
presents our conclusions.

Throughout this paper spectral indices are written $S_{\nu} \propto
\nu^{-\alpha}$ and the values $H_0 = 50$ km s$^{-1}$ Mpc$^{-1}$ and $q_0 = 0$
have been adopted.

\section{The Sample}
The 1-Jy sample of BL Lacs is presently the only sizeable, complete sample of
radio bright BL Lacs. It includes 34 objects with radio flux $> 1$ Jy at 5 GHz
and $\alpha_{\rm r} \le 0.5$ (Stickel et al. 1991). All 1-Jy BL Lacs have been
studied in detail in the radio and optical bands; all objects have also soft
X-ray data, primarily from \rosat.

We selected for \sax observations all 1-Jy BL Lacs with $0.1 - 10$ keV X-ray
flux larger than $2 \times 10^{-12}$ erg cm$^{-2}$ s$^{-1}$ (estimated from an
extrapolation of the single power-law fits derived for these objects from
\rosat~data; Urry et al. 1996). This included twenty 1-Jy BL Lacs (or $\sim
60$ per cent of the sample). Observing time was granted for 11 of them, while
three more sources have been included in other
\sax programs (MKN 501: Pian et al. 1998; S5 0716+714: Giommi et al. 1999; PKS
0537$-$441: Pian et al. 2002). We present here the results obtained for the
four objects observed in Cycle 2, all LBL. The object list and basic
characteristics are given in Table \ref{sample}, which presents the source
name, position, redshift, $R$ magnitude, 5 GHz radio flux, and Galactic \nh.

\begin{table*}
\caption{Sample Properties.\label{sample}}
\begin{tabular}{lrrllrl}
\hline
Name & RA(J2000) & Dec(J2000)&~~~$z$ & R$_{\rm mag}^a$ & F$_{\rm 5GHz}$&Galactic N$_{\rm H}$ \\
              &            &             &    &   &(Jy) &($10^{20}$ cm$^{-2}$) \\
\hline
AO 0235$+$164   & 02 38 38.9 & $+$16 36 59 &0.940 & 16.5 & 2.9 & 7.60{\footnotesize 
$^{b} $} \\
OQ 530          & 14 19 46.6 & $+$54 23 15 &0.152 & 14.5 & 1.1 & 1.20{\footnotesize 
$^{c} $} \\   
S5 1803$+$784   & 18 00 45.7 & $+$78 28 04 &0.684 & 16.5 & 2.6 & 3.65{\footnotesize 
$^{d} $} \\
3C 371          & 18 06 50.6 & $+$69 49 28 &0.051 & 14.0 & 2.3 & 4.38{\footnotesize 
$^{d} $} \\ 
\hline
\multicolumn{6}{l}{\footnotesize{$^a$ Mean R magnitude from Heidt \& Wagner
(1996).}} \\
\multicolumn{6}{l}{\footnotesize{$^b$ From Elvis et al. (1989).}} \\
\multicolumn{6}{l}{\footnotesize{$^c$ Dickey \& Lockman (1990).}} \\
\multicolumn{6}{l}{\footnotesize{$^d$ From Murphy et al. (1996).}} \\
\end{tabular}
\end{table*}

\section{Observations and Data Analysis}
%
A complete description of the \sax mission is given by Boella \etal (1997a).
The relevant instruments for our observations are the coaligned Narrow Field
Instruments (NFI), which include one Low Energy Concentrator Spectrometer
(LECS; Parmar \etal 1997) sensitive in the 0.1 -- 10 keV band; two (three 
before 1997 May) identical
Medium Energy Concentrator Spectrometers (MECS; Boella \etal 1997b), covering
the 1.5 -- 10 keV band; and the Phoswich Detector System (PDS; Frontera \etal
1997), coaligned with the LECS and the MECS. The PDS instrument is made up of
four units, and was operated in collimator rocking mode, with a pair of units
pointing at the source and the other pair pointing at the background, the two
pairs switching on and off source every 96 seconds. The net source spectra
have been obtained by subtracting the `off' to the `on' counts. 
A journal of
the observations is given in Table \ref{saxobs}. 

The data analysis was based on the linearized, cleaned event files obtained
from the \sax Science Data Center (SDC) on-line archive (Giommi \& Fiore
1997). The data from the two MECS instruments were merged in one single event
file by the SDC, based on sky coordinates.  The event file was then screened
with a time filter given by SDC to exclude those intervals related to events
without attitude solution (i.e., conversion from detector to sky coordinates;
see Fiore et al. 1999). This was done to avoid an artificial decrease in the
flux. As recommended by the SDC, the channels 1--10 and above 4 keV for the
LECS and 0--36 and 220--256 for the MECS were excluded from the spectral
analysis, due to residual calibration uncertainties. None of our sources was
detected by the PDS.

Spectra and lightcurves were extracted using the standard radius of 4
arcmin for the MECS. As regards the LECS, the recommended value of 8 arcmin
was used only for S5 1803+784, while 6 and 4 arcmin were used
for AO 0235+164 and OQ 530, respectively, due to their very low signal-to-noise ratio (S/N). 
A 4 arcmin radius was used for 3C 371, due to the presence of 
a serendipitous source at $\sim 7$ arcmin (see below).

The spectral analysis was performed using the matrices and blank-sky
background files released in 1998 November by the SDC, with the blank-sky
files extracted in the same coordinate frame as the source file, as described
in Paper I. Because of the importance of the band below 1 keV
to assess the presence of extra absorption or soft excess (indicative of a
low-energy steepening of the spectrum), we have also checked the LECS data for 
differences in
the cosmic background between local and blank-sky field observations,
comparing spectra extracted from the same areas on the detector (namely, two
circular regions outside the 10 arcmin radius central region, located at
the opposite corners with respect to the two on-board radioactive calibration
sources). No significant differences were found.

\begin{figure}
\centerline{\psfig{figure=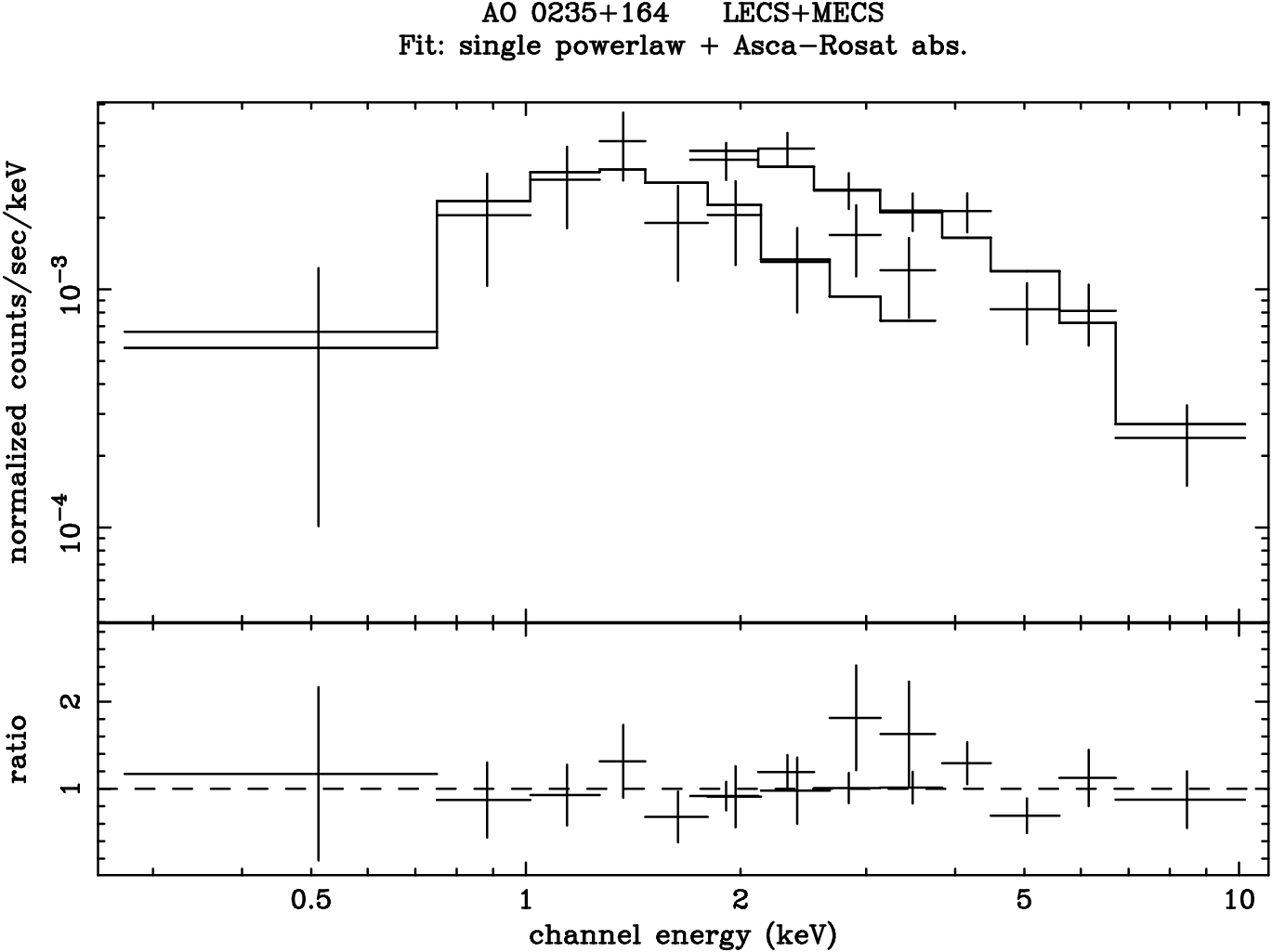,width=9cm,angle=-90}}
\medskip
\centerline{\psfig{figure=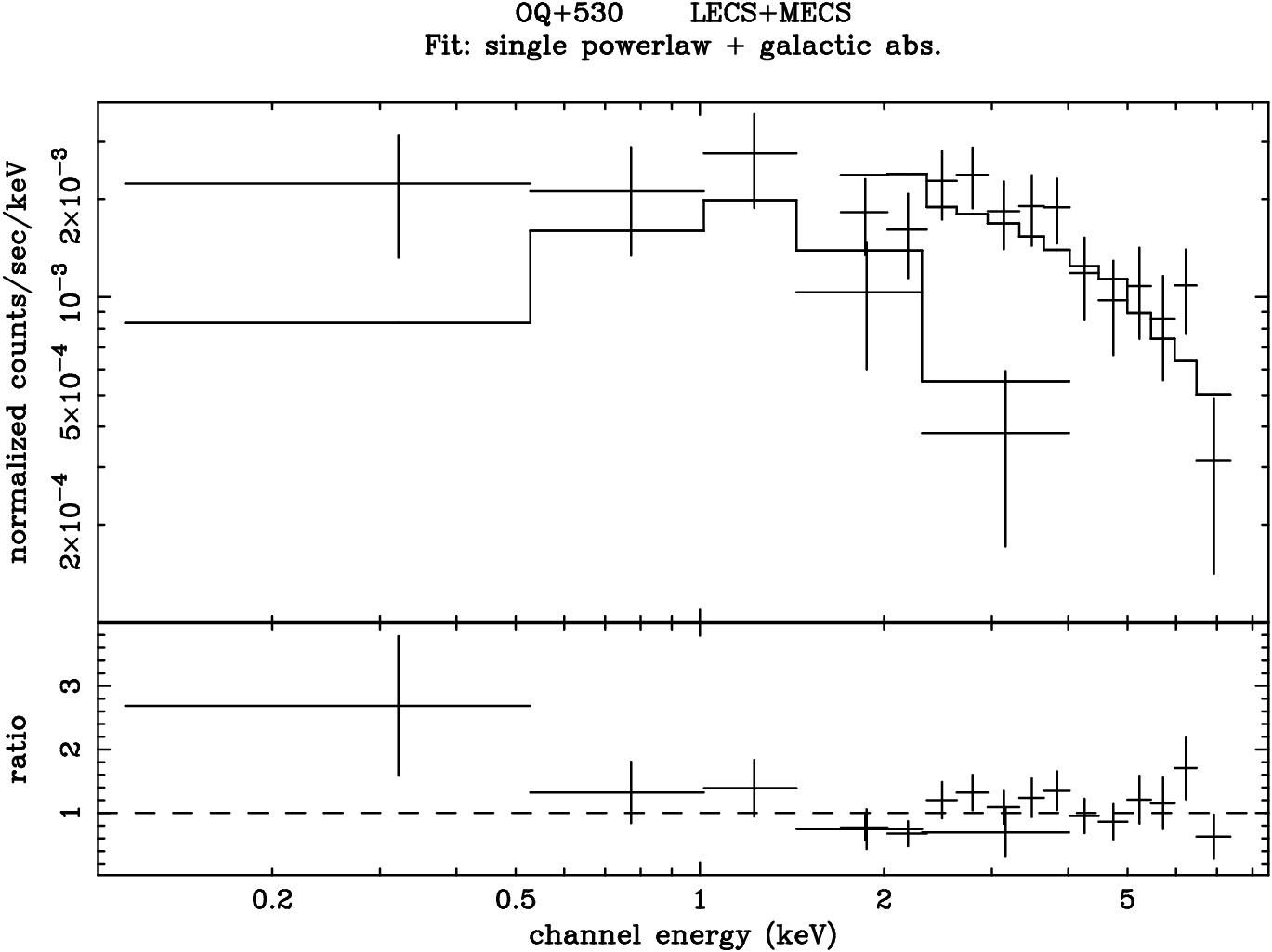,width=9cm,angle=-90}}
\medskip
\centerline{\psfig{figure=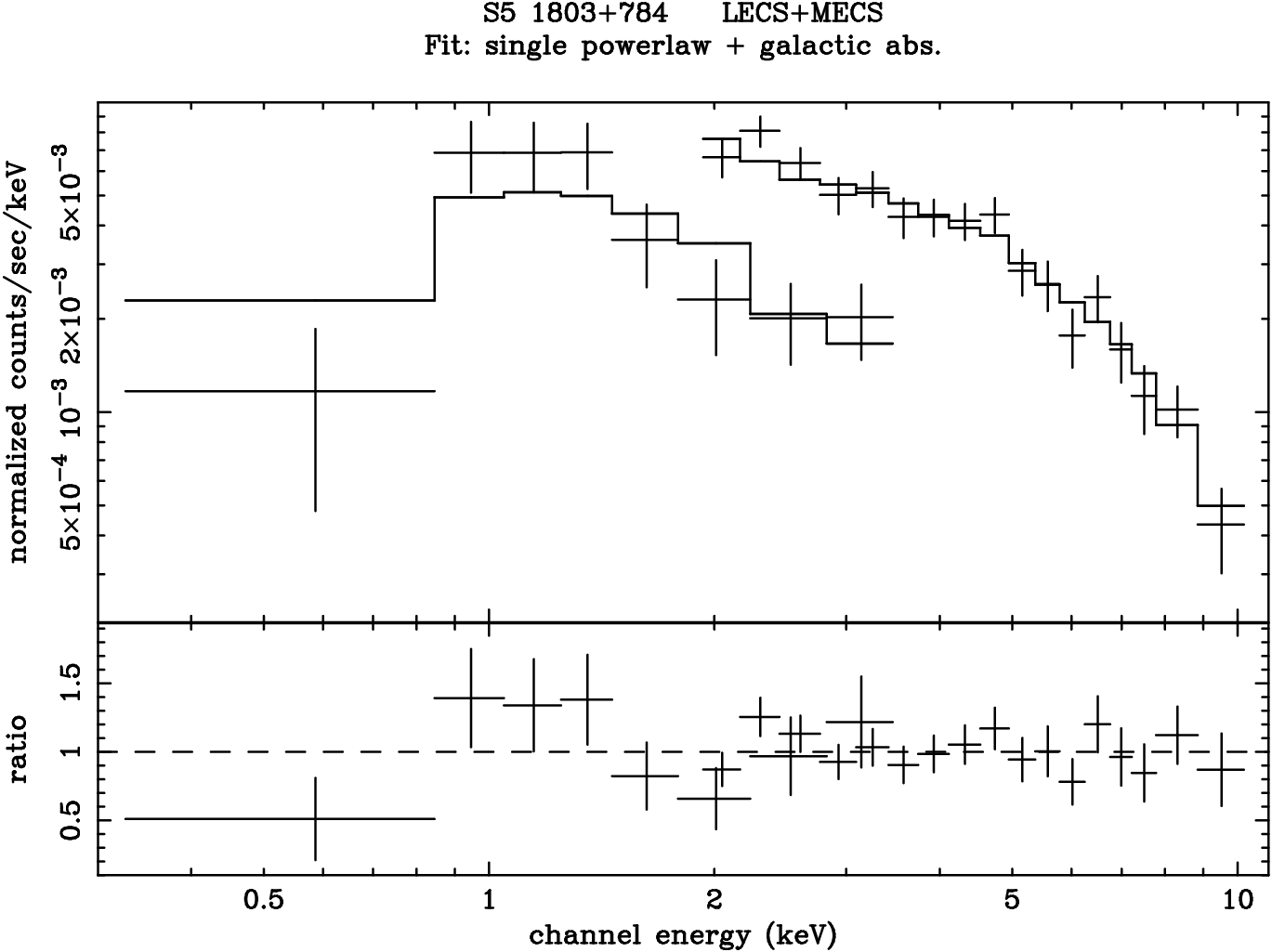,width=9cm,angle=-90}}
\end{figure}
\begin{figure}
\centerline{\psfig{figure=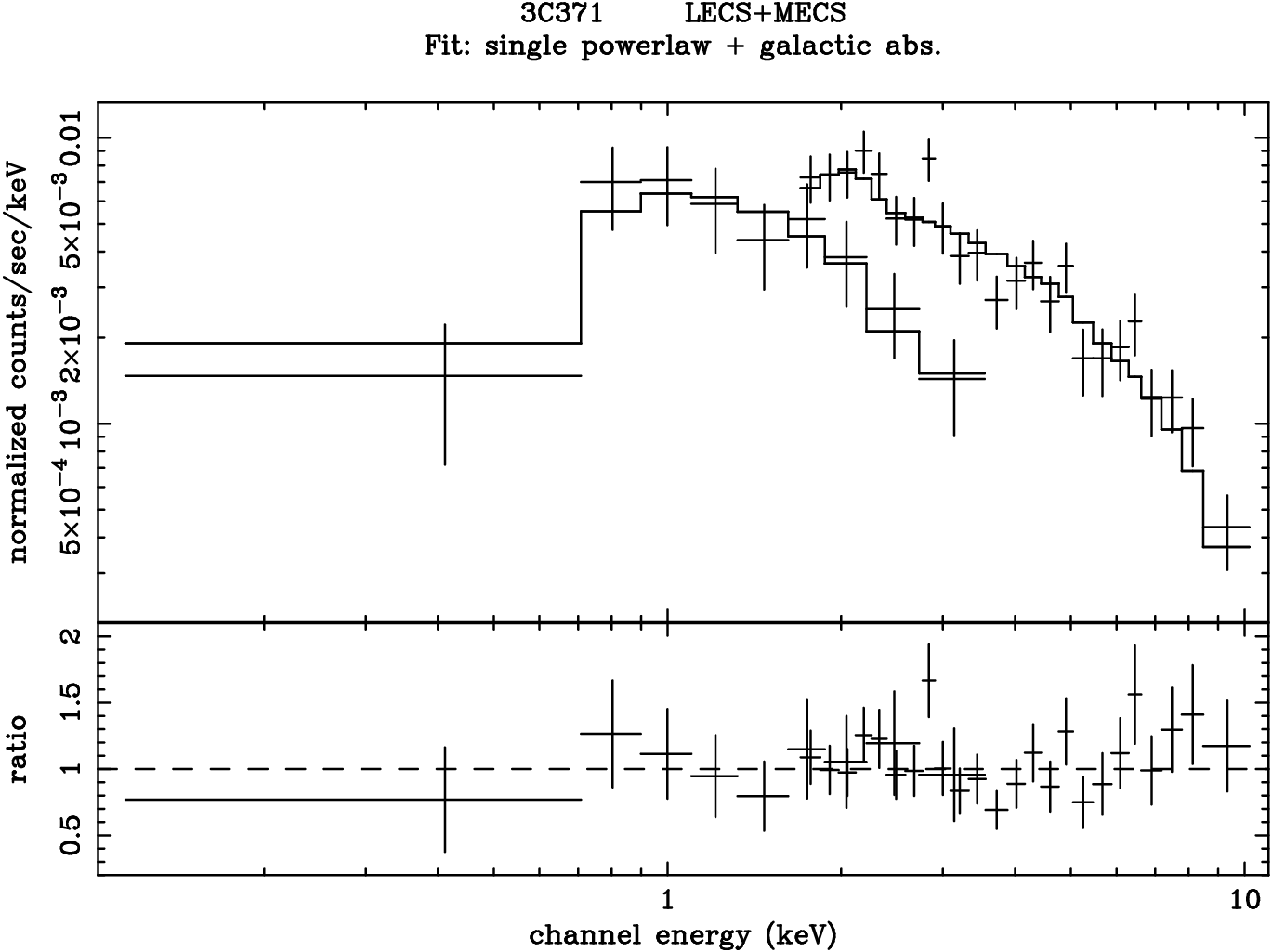,width=9cm,angle=-90}}
\medskip
\caption{\sax data and fitted spectra for our sources, and ratio of data to
fit. Data are from the LECS and MECS instruments. The data are fitted with a
single power-law model with Galactic absorption, apart from AO 0235+164, for
which the absorption includes also the contribution of an intervening system
at $z = 0.524$ (Madejski et al. 1996).\label{saxspectra}}
\end{figure}
\setcounter{figure}{1}

\begin{table*}
\caption{BeppoSAX Journal of observations.\label{saxobs}}
\begin{tabular}{lrcrcc}
\hline
Name &LECS& LECS &MECS & MECS & Observing date\\
 &Exp. & Count rate$^{a}$ &Exp.& Count rate$^{a}$   & \\
 &(s)  & (cts/s)          &(s) & (cts/s) & \\
\hline
AO 0235+164   &    24575&$0.006\pm0.002 $&    29071& $0.011\pm0.001 $ & 1999 
Jan 28 \\
OQ  530       &    16312&$0.006\pm0.002 $&    39680& $0.008\pm0.001 $ & 1999 
Feb 12-13 \\
S5 1803+784   &    18359&$0.013\pm0.002 $&    40303& $0.028\pm0.001 $ & 1998 
Sep 28 \\
3C 371        &    13090&$0.018\pm0.002 $&    33760& $0.024\pm0.001 $ & 1998 
Sep 22-23 \\
\hline
\multicolumn{6}{l}{\footnotesize $^a$ Net count rate full band.} \\
\end{tabular}
\end{table*}

Using the software package XRONOS we looked for time variability in every
observation, binning the data in intervals from 500 to 3600 s, with null 
results.
 
\section{Spectral Fits}\label{fitssax}

Spectral analysis has been performed with the XSPEC 10.00 package, using
the response matrices released by the SDC in 1998. Using the program GRPPHA, the
spectra were rebinned with more than 20 counts in every new bin and 
using the rebinning files provided by SDC. Various checks using different
rebinning strategies have shown that our results are independent of the
adopted rebinning within the uncertainties. The data were analyzed applying 
the Gehrels
statistical weight (Gehrels 1986) in case the resulting net counts were below 20 
(typically
12-15 in the low energy band of LECS). The LECS/MECS normalization
factor was left free to vary in the range 0.65--1.0, as suggested by SDC (see
Fiore et al. 1999). The X-ray spectra of our 
sources are shown in Fig. \ref{saxspectra}. 

\subsection{Single Power-law Fits}

At first, we fitted the combined LECS and MECS data with a single power-law
model with Galactic and free absorption. The absorbing column was
parameterized in terms of N$_{\rm H}$, the HI column density, with heavier
elements fixed at solar abundances and cross sections taken from Morrison and
McCammon (1983). The \nh~value for AO 0235+164 was also fixed at a value
larger than the Galactic one to take into account absorption by an intervening
galaxy (Madejski et al. 1996). \nh~was also set free to vary for all sources to 
check for internal absorption and/or indications of a ``soft-excess.''


\begin{table*}
\begin{center}
\caption{Single power-law fits, LECS + MECS$^a$.\label{saxfits}}
\begin{tabular}{lllllllll}
\hline
Name  & N$_H$ & $\alpha_{x}$ & $F_{1keV}$ & $F_{[2-10]}$ & $F_{[0.1-2.4]}$ & Norm    & $\chi^2_{\nu}$/dof &  
$F$-test, notes \\
     & ($10^{20}$ cm$^{-2}$) & & ($\mu$Jy) & (c.g.s.) & (c.g.s.) & (L/M) & &                        
fixed-free N$_{\rm H}$\\
\hline
AO 0235$+$164   & 7.6 fixed      & $0.79^{+0.24}_{-0.24}$ & $0.16^{+0.06}_{-0.05}$ & 8.54e-13 & 
1.08e-12 & 0.70 & 0.97/14 &   \\
              & $37^{+85}_{-33}$ & $1.01^{+0.47}_{-0.41}$ & $0.22^{+0.16}_{-0.08}$ & 8.43e-13 & 
1.71e-12 & 0.82 & 0.78/13 & 94 per cent \\
              & 28$^b$ fixed     & $0.96^{+0.27}_{-0.26}$ & $0.20^{+0.08}_{-0.06}$ & 8.42e-13 &
1.51e-12 & 0.79 & 0.74/14 & N$_H$ at ASCA and \rosat~values
\vspace*{1mm}\\
\hline
\vspace*{-3mm} \\
OQ    530       & 1.20 fixed        & $0.55^{+0.27}_{-0.32}$ & $0.09^{+0.03}_{-0.03}$ & 6.74e-13 & 
5.24e-13 & 0.74 & 1.12/15 &   \\
              & $0.0 (<7.38)$       & $0.55^{+0.32}_{-0.39}$ & $0.09^{+0.03}_{-0.03}$ & 6.75e-13 & 
5.24e-13 & 0.71 & 1.09/14  & 74 per cent 
\vspace*{1mm}\\
\hline
\vspace*{-3mm} \\
S5 1803$+$784   & 3.65 fixed       & $0.45^{+0.12}_{-0.12}$ & $0.24^{+0.04}_{-0.04}$ & 2.21e-12 & 
1.41e-12 & 0.67 & 0.97/22 &   \\
              & $10 (<33)$         & $0.51^{+0.20}_{-0.20}$ &
              $0.26^{+0.07}_{-0.05}$ & 2.20e-12 & 
1.57e-12 & 0.67 & 0.97/21 & 65 per cent 
\vspace*{1mm}\\
\hline
\vspace*{-3mm} \\
3C 371        & 4.38 fixed           & $0.72^{+0.16}_{-0.16}$ & $0.30^{+0.07}_{-0.07}$ & 1.76e-12 & 
1.93e-12 & 0.72 & 0.86/30 &   \\
              & $5.6^{+19}_{-4.6}$   & $0.74^{+0.24}_{-0.24}$ & $0.30^{+0.08}_{-0.08}$ & 1.76e-12 & 
1.99e-12 & 0.73 & 0.89/29 & 30 per cent
\vspace*{1mm}\\
\hline
\vspace*{-3mm} \\ \multicolumn{9}{l}{\footnotesize{$^a$Errors are 
at $90$ per cent confidence level for one (with
fixed N$_{\rm H}$) and two parameters of interest.}} \\
\multicolumn{9}{l}{\footnotesize{$^b$Absorption including the
contribution of an intervening system at z=0.524 (Madejski et al. 1996).}} \\
\end{tabular}
\end{center}
\end{table*}

Our results are presented in Table \ref{saxfits}, which gives the name of the
source in column (1), \nh~in column (2), the energy index $\alpha_{\rm x}$ in
column (3), the 1 keV flux in $\mu$Jy in column (4), the unabsorbed $2 - 10$
keV and $0.1 - 2.4$ keV fluxes in columns (5)-(6), the LECS/MECS normalization
in column (7), the reduced chi-square and number of degrees of freedom,
$\chi^2_{\nu}/$(dof) in column (8), and the $F$--test probability that the
decrease in $\chi^2$ due to the addition of a new parameter (free \nh) is
significant in column (9). The errors quoted on the fit parameters are the 90
per cent uncertainties for one and two interesting parameters, for Galactic
and free \nh~respectively. The errors on the 1 keV flux reflect the
statistical errors only and not the model uncertainties.
 
Two results are immediately apparent from Table \ref{saxfits}. First, the
fitted \nh~values do not agree with the Galactic ones for AO 0235+164 (as
expected from previous ASCA and {\it ROSAT} observations; see Sect. 4.2.1) and
possibly OQ 530. An $F$-test shows that the addition of \nh~as a free
parameter results in a (marginally significant) improvement in the $\chi^2$
values only for AO 0235+164 (column 9 of Table \ref{saxfits}). Second, the
fitted energy indices are flat, \ax~$< 1$, with $\langle \alpha_{\rm x}
\rangle = 0.67\pm0.11$ and a weighted mean equal to $0.59\pm0.09$.

One of our sources, OQ 530, appears to show a low-energy excess, as
illustrated by Fig. \ref{saxspectra} and by the fact that the best fit \nh~in
Table \ref{saxfits} is below the Galactic value (and actually consistent with
zero). We then fitted a broken power-law model to the data. The results are
presented in Table \ref{saxbkn}, which gives the name of the source in column
(1), \nh~in column (2), the soft energy index $\alpha_{\rm S}$ in column (3),
the hard energy index $\alpha_{\rm H}$ in column (4), the break energy in
column (5), the 1 keV flux in $\mu$Jy in column (6), the unabsorbed $2 - 10$
keV and $0.1 - 2.4$ keV fluxes in columns (7)-(8), the LECS/MECS normalization
in column (9), the reduced chi-squared and number of degrees of freedom,
$\chi^2_{\nu}/$(dof), in column (10), and the $F$--test probability that the
decrease in $\chi^2$ due to the addition of two parameters (from a single
power-law fit to a broken power-law fit) is significant in column
(11). Although the fit is improved by using a double power-law model, an
$F$--test shows that the improvement is more suggestive than significant, with
a probability $\sim 90$ per cent. The best-fit spectrum, however, points in
the direction of a flatter component emerging at higher energies, as was the
case for four sources in Paper I. In fact, the spectrum is concave with a
large spectral change $\langle \alpha_{\rm S} - \alpha_{\rm H} \rangle \sim
0.9$, and an energy break around $E \sim 2$ keV. Evidence for a concave
spectrum comes also from the shape of the ratio of the data to the single
power-law fits, shown in Fig. \ref{saxspectra}.

\begin{table*}
\caption{Broken power-law fits, LECS + MECS$^a$.\label{saxbkn}}
\begin{tabular}{lllllllllll}
\vspace*{-1mm}\\
\hline
\vspace*{-3mm} \\
Name    & N$_{\rm H}$     & $\alpha_S$  & $\alpha_H$  &  E$_{break}$ & $F_{1keV}$ &  $F_{[2-10]}$  & 
$F_{[0.1-2.4]}$ & L/M & $\chi^2_{\nu}$/dof & $F$-test$^b$ \\
    & ($10^{20}$ cm$^{-2}$) &       &           &  (keV) &  ($\mu$Jy) & (c.g.s.)  & (c.g.s.)  &    &     &         
\vspace*{1mm}\\
\hline
\vspace*{-3mm} \\
OQ  530    &  1.20 fixed    & $1.3^{+4.0}_{-0.7}$ &
$0.4^{+0.4}_{-0.4}$ & $1.8^{+1.1}_{-1.5}$
  & $0.12^{+0.03}_{-0.07}$        & 7.31e-13  & 1.10e-12 & 0.66 & 0.91/13
  & 90 per cent 
\vspace*{1mm}\\
\hline
\vspace*{-3mm} \\ \multicolumn{11}{l}{\footnotesize{$^a$Errors are at 
90 per cent confidence level for two parameters of
interest.}} \\ \multicolumn{11}{l}{\footnotesize{$^b$The values of the
$F$--test refer to the comparison with a single power-law model with Galactic
column density.}}\\
\end{tabular}
\end{table*}

\subsection{Notes on individual sources}

\subsubsection{AO 0235+164}

This source is in a crowded field, with three absorbing systems
at $z= 0.524$, 0.851, and 0.94 in a 10x10 arcsec region and a probable
Seyfert 1 galaxy at less than 5 arcsec (see Burbidge et al. 1996). \rosat~and
ASCA spectra show absorption above the Galactic value, which our data confirm,
likely due to the intervening absorption system at $z=0.524$ (Madejski et
al. 1996). 
A fit with free absorption, in fact, gives an \nh~value slightly higher
than the corresponding ASCA one (3.7$\times 10^{21}$ cm$^{-2}$ vs. 2.8
$\times 10^{21}$ cm$^{-2}$) but still consistent within the errors.
Since the ASCA data provide a better constraint on the column density,
we have also fitted our data with total \nh~fixed at the ASCA value. 
The \sax~X-ray spectral index and flux are consistent with the 
values of Madejski et al. (1996), under the same assumptions. 

Given the \sax~point spread function, it is nearly impossible to 
assess the likelihood of a contribution from the Seyfert galaxy to the X-ray
flux of our source. However, the MECS profile is the one expected from a
point-like source, with no hints of extension. Moreover, using its observed
optical magnitude ($\sim 20.5$) and assuming an optical-X-ray effective
spectral index typical of Seyfert 1s ($\alpha_{\rm ox} \sim 1.2$), one can
estimate an X-ray flux from this nearby source roughly a factor of 10 smaller
than that observed for AO 0235+164.

\subsubsection{OQ 530} 

This source shows some evidence of a steeper spectrum at low energies,
but the $F$-test significance is only marginal, due to the low statistics 
and narrow range affected.
Ta\-glia\-ferri et al. (2003) have also presented evidence for a low-energy
excess, with a very steep $\alpha_{\rm S} \sim 6.7$ and $\alpha_{\rm H} =
0.4\pm0.2$ (2000 March 3-4) and $\alpha_{\rm H} = 0.75\pm0.20$ (2000 March
26-27). In their case the $F$-test probability for improvement for the double 
power-law
model is $>99.9$ per cent and $\sim 80$ per cent, respectively. The fluxes
for the two observations are similar, $f({\rm 2 - 10 keV}) \sim 10^{-12}$
erg cm$^{-2}$ s$^{-1}$, $\sim 50$ per cent larger than what we find. 


\subsubsection{3C 371} 

A serendipitous source identified as RIXOS F272$\_$023, a Seyfert 1.8 galaxy
at $z=0.096$, is present in the LECS and MECS images, at a distance of
$\sim 7$ arcmin (Puchnarewicz et al. 1997). 
The extraction of LECS data was then done by using a radius of 4 arcmin.


\section{\rosat~PSPC data}

In order to compare our results with previous (soft) X-ray observations and
especially to take advantage of the higher resolution and collecting area at
low energies, we used data from the \rosat~Position Sensitive Proportional
Counter (PSPC). The 1-Jy BL Lac \rosat~data had been originally published by
Urry et al. (1996). In order to ensure a uniform procedure for the whole
sample, we have re-analyzed all \rosat~data, obtaining results consistent
within the errors with those already published. 

\begin{table*}
\caption{\rosat~journal of observations.\label{rosatobs}}
\begin{tabular}{lrcc}
\hline
Name &  Exposure  &  Full band net countrate & Observing Date \\
     &   (s)    &    (cts/s) &    \\
\hline
AO 0235+164  &\dig17949 & $0.170\pm0.003$ & 1993 Jul 21 -- Aug 15\\
OQ 530      &\dig11474 & $0.223\pm0.005$ & 1990 Jul 19 -- 23\\
S5 1803+784  &\dig6781  & $0.081\pm0.004$ & 1992 Apr 7\\
             &\dig2773  & $0.105\pm0.008$ & 1992 Jul 25\\
             &\dig2827  & $0.108\pm0.008$ & 1992 Dec 10\\
3C 371       &\dig10348 & $0.127\pm0.004$ & 1992 Apr 9\\
\hline
\end{tabular}
\end{table*}

The journal of the \rosat~observations is given in Table \ref{rosatobs}. 
The basic event
files from the archive have been corrected for gain variations on the detector
surface with the program PCSASSCORR in FTOOLS, when not already done by the
Standard Reduction process (SASS version 7\_8 and later, M. Corcoran, private
communication). Since all the sources were \rosat~targets a standard 
extraction radius of $3^{\prime}$ ($2.5^{\prime}$
when serendipitous sources were present in the field or when the source was
particularly weak) was used, to avoid the possible loss of soft photons due to
the ghost imaging effect. We have used the appropriate response matrices for
the different gain levels of the PSPC B detector before and after 14 Oct. 1991
(gain1 and gain2, respectively). The background has been evaluated in two
circular regions (of radius $\sim 20-30$ pixels) away from the central region
and from other serendipitous sources, but inside the central rib of the
detector. The spectra have been rebinned (using GRPPHA) with more than 20
counts in every new bin, to validate the use of \chisq~statistics. 
Channel 1-11 and 212-256 have been excluded from the analysis, due to 
calibration uncertainties. 

\begin{table*}
\begin{center}
\caption{\rosat~PSPC,  single power-law fits$^a$.\label{rosatfits}}
\begin{tabular}{lllllll}
\vspace*{-1mm}\\
\hline
\vspace*{-2mm} \\
Name  & N$_{\rm H}$ & $\alpha_{x}$ & $F_{1keV}$ & $F_{[0.1-2.4]}$ & $\chi^2_{\nu}$/dof& Observing Date\\
      & ($10^{20}$ cm$^{-2}$) & & ($\mu$Jy) & (erg cm$^{-2}$ s$^{-1}$) &  &   
\vspace*{1mm}\\
\hline
\vspace*{-3mm} \\
AO 0235+164    & 7.60 fixed             & $0.50\pm0.08$ & $0.64\pm0.02$ & 3.84e-12 & 3.22/30 & 1993 Jul 21 -- Aug 15 \\
               & $33^{+12}_{-10}$       & $2.1\pm0.7$   & $1.4^{+0.4}_{-0.3}$ & 4.45e-11 & 0.79/29  & \\
               & 28$^b$ fixed               & $1.88\pm0.11$ & $1.21\pm0.04$   & 2.37e-11   &  0.79/30  &    
%
%
\vspace*{1mm}\\
\hline
\vspace*{-3mm} \\
OQ 530$^c$    & 1.20 fixed             & $1.07\pm0.05$ & $0.30\pm0.02$ & 2.47e-12 & 1.32/76 & 1990 Jul 19 -- 23\\  
               & $1.2^{+0.4}_{-0.3}$    & $1.1\pm0.2$   & $0.30\pm0.02$ & 2.50e-12 & 1.34/75 &
\vspace*{1mm}\\
\hline
\vspace*{-3mm} \\
S5 1803$+$784 & 3.65 fixed             & $0.84\pm0.15$      & $0.23\pm0.02$   & 1.61e-12 & 0.89/21 & 1992 Apr 7\\
              & $4.2^{+1.9}_{-1.6}$    & $1.0\pm0.4$        & $0.24\pm0.02$   & 1.79e-12 & 0.91/20 &         
\vspace*{3mm}\\
              & 3.65 fixed             & $1.2\pm0.2$        & $0.27\pm0.03$   & 2.40e-12 & 0.68/10 & 1992 Jul 25\\
              & $3.5^{+2.4}_{-2.0}$    & $1.1\pm0.7$        & $0.27\pm0.04$   & 2.28e-12 & 0.76/9  &                       
\vspace*{3mm}\\
              & 3.65 fixed             & $1.2\pm0.2$        & $0.26\pm0.03$   & 2.38e-12 & 1.32/10 & 1992 Dec 10\\ 
              & $4.6^{+2.8}_{-2.4}$    & $1.5\pm0.8$        & $0.26\pm0.04$   & 3.20e-12 & 1.40/9  &         
\vspace*{1mm}\\
\hline
\vspace*{-3mm} \\
3C 371        & 4.38 fixed            & $1.21\pm0.10$      & $0.35\pm0.02$   & 3.20e-12 & 1.36/25 & 1992 Apr 9\\
              & $2.7^{+0.9}_{-0.8}$    & $0.8\pm0.3$        & $0.33\pm0.02$   & 2.21e-12 & 0.85/24 & 
\vspace*{1mm}\\
\hline
\vspace*{-3mm} \\ 
\multicolumn{7}{l}{\footnotesize{$^a$Errors are at 90 per
cent confidence level for one (with fixed N$_{\rm H}$) and two parameters of
interest.}} \\ 
\multicolumn{7}{l}{\footnotesize{$^b$Absorption including the
contribution of an intervening system at z=0.524 (Madejski et al. 1996).}} \\
\multicolumn{7}{l}{\footnotesize{$^c$PSPC C}}
\end{tabular}
\end{center}
\end{table*}

\begin{table*}
\caption{Broken power-law fits, \rosat$^a$.\label{rosatbkn}}
\begin{tabular}{lllllllll}
\vspace*{-1mm}\\
\hline
\vspace*{-3mm} \\
Name    & N$_{\rm H}$     & $\alpha_S$  & $\alpha_H$  &  E$_{break}$ & $F_{1keV}$  & 
$F_{[0.1-2.4]}$ & $\chi^2_{\nu}$/dof & $F$-test$^b$ \\
    & ($10^{20}$ cm$^{-2}$) &       &           &  (keV) & ($\mu$Jy)  & (c.g.s.)  &     &         
\vspace*{1mm}\\
\hline
\vspace*{-3mm} \\
3C 371    &  4.38 fixed    & $3.0^{+2.4}_{-2.2}$ & $0.9\pm0.2$ & $0.34^{+0.64}_{-0.06}$
  & $0.3\pm0.1$            & 5.60e-12  & 0.88/23  & 99.7 per cent 
\vspace*{1mm}\\
\hline
\vspace*{-3mm} \\ \multicolumn{9}{l}{\footnotesize{$^a$Errors are at 90 per
cent confidence level for two parameters of interest.}} \\
\multicolumn{9}{l}{\footnotesize{$^b$The values of the $F$--test refer to the
comparison with a single power-law model with Galactic column density.}}\\
\end{tabular}
\end{table*}

As for the \sax~data, we fitted the \rosat~PSPC data with a single power-law
model with Galactic and free absorption. Our results are presented in Table
\ref{rosatfits}, which gives the name of the source in column (1), \nh~in
column (2), the energy index $\alpha_{\rm x}$ in column (3), the 1 keV flux in
$\mu$Jy in column (4), the unabsorbed $0.1 - 2.4$ keV flux in column (5), the
reduced chi-squared and number of degrees of freedom, $\chi^2_{\nu}/$(dof) in
column (6), and the observing date in column (7). 

Table \ref{rosatfits} shows that the fitted \nh~values are consistent with the
Galactic ones for OQ 530 and S5 1803+784. AO 0235+164 shows evidence of excess
absorption, as was the case for our \sax~data, while 3C 371 displays a
soft-excess, suggestive of a steeper soft component. A broken power-law fit,
in fact, improves the fit significantly (99.7 per cent), as shown in Table
\ref{rosatbkn}. The spectrum is obviously concave with quite a large spectral
change, with $\langle \alpha_{\rm S} - \alpha_{\rm H} \rangle \sim 2$, and an
energy break around $E \sim 0.3$ keV.

The fitted energy indices are relatively steep. The mean \rosat~value for our
sources is $\langle \alpha_{\rm x} \rangle = 1.34\pm0.18$, while the weighted
mean is $\langle \alpha_{\rm x} \rangle = 1.21\pm0.04$.

\begin{figure}
\centerline{\psfig{figure=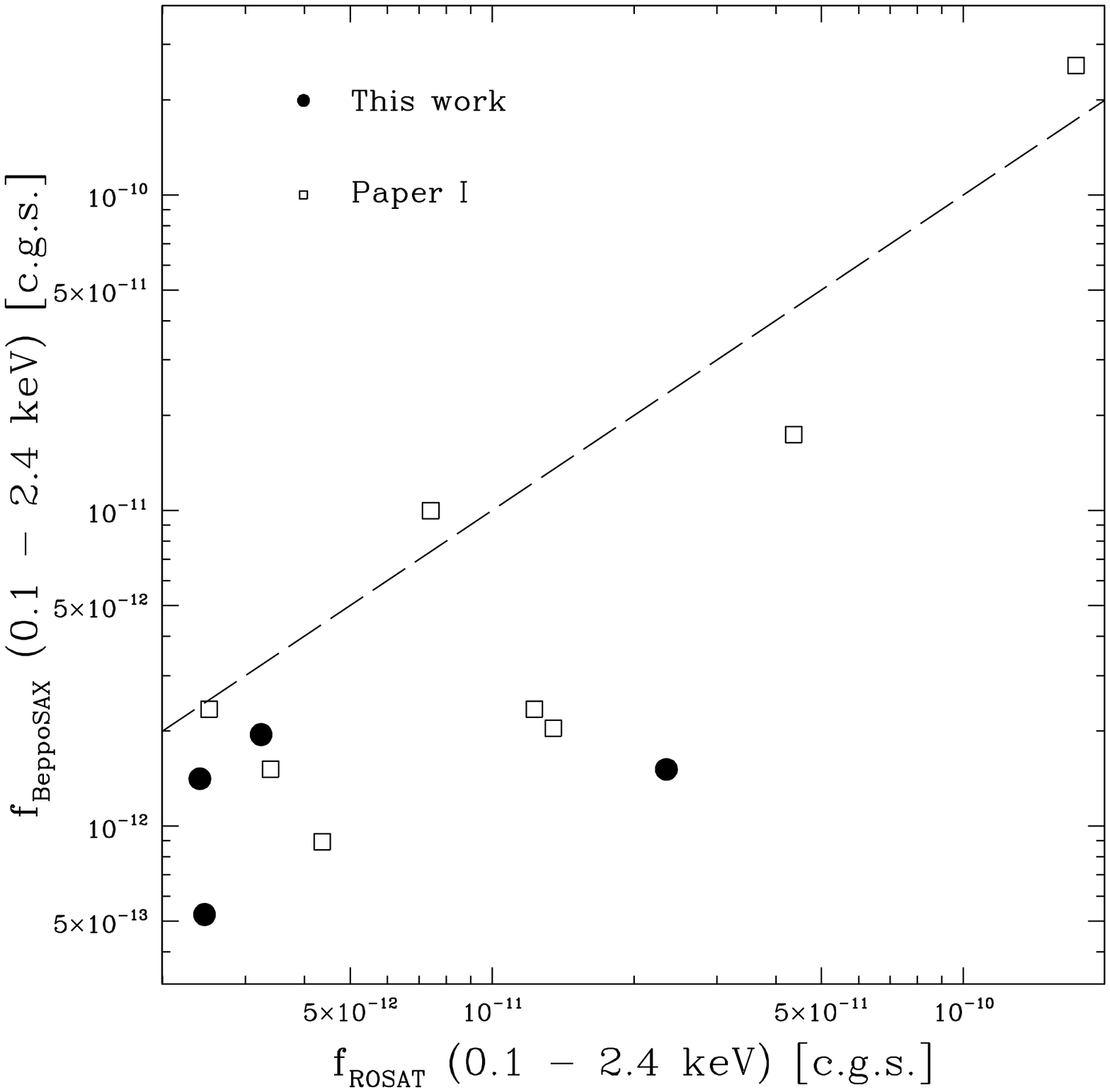,width=9cm}}
\caption{The $0.1 - 2.4$ keV X-ray flux from our \sax data vs. the 
corresponding \rosat~flux (filled points). The dashed line represents the 
locus of $f_{\sax}=
f_{\rosat}$. For S5 1803+784, which has multiple \rosat~observations, we took
the observation with the largest X-ray flux (1992 July). Open squares
represent the BL Lacs studied in Paper I.\label{fluxes}}
\end{figure}

\begin{figure}
\centerline{\psfig{figure=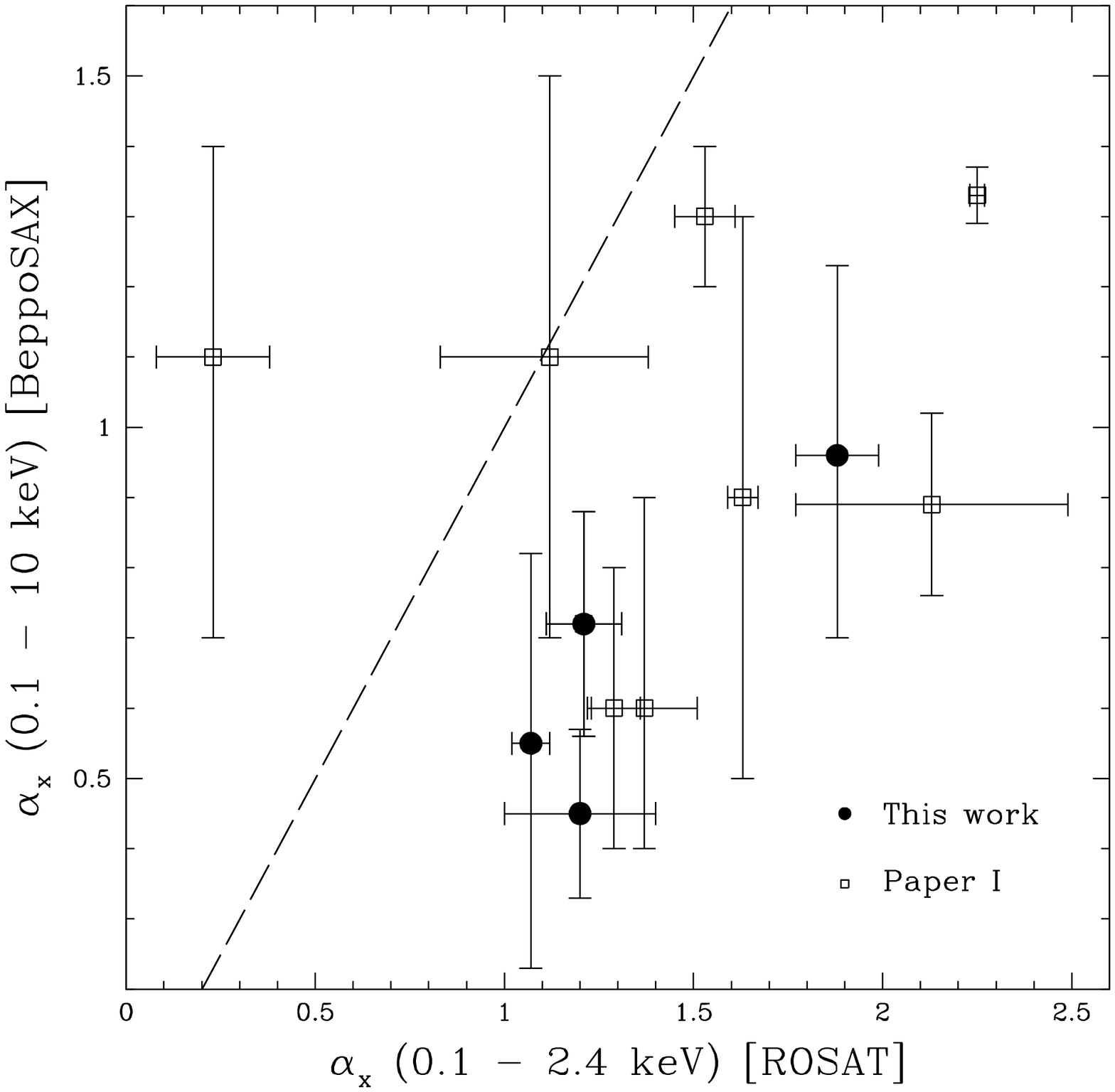,width=9cm}}
\caption{The \sax spectral index ($0.1-10$ keV) vs. the \rosat~spectral index
($0.1-2.4$ keV) for our sources (filled points). The dashed line represents
the locus of $\alpha_{\rm x}(\sax)= \alpha_{\rm x}(\rosat)$. For S5 1803+784,
which has multiple \rosat~observations, we took the observation with the
largest X-ray flux (1992 July). Open squares represent the BL Lacs studied in
Paper I.\label{alpha_sax_rosat}}
\end{figure}

\subsection{Notes on individual sources}

\subsubsection{AO 0235+164}

Nine \rosat~observations are available, spaced by about $\sim3$ days, between
July 21 and August 15 1993. The source shows some interesting variability,
but with  no evident spectral variations, as the hardness ratio is roughly
constant. Therefore, the data have been summed together (see Madejski et
al. 1996 for a discussion of the single observations). 
 
\subsubsection{OQ 530} 

The \rosat~observations are relatively old and done with the PSPC C
detector. The spectrum has many "wiggles", which suggest that perhaps the
calibration is not optimal. The resulting reduced $\chi^2$ is not very good,
but alternative models (free \nh, broken power-law) do not improve the fit.

\subsubsection{S5 1803+784} 
Three \rosat~observations are available. The best-fit \nh~agrees with the
Galactic value. The source displays interesting spectral and intensity
variations, with a ``steeper when brighter'' behaviour, which suggests
properties typical of ``intermediate'' BL Lacs. These are possibly associated
with a shift of the synchrotron peak or a hardening of the injected electron
spectrum in higher states, so that synchrotron emission becomes dominant in the
soft X-ray band.

\subsection{Comparison between \sax and \rosat~results}

Figure \ref{fluxes} shows the $0.1 - 2.4$ keV \sax flux versus the 
corresponding \rosat~
flux. Our sources display mild X-ray variability: the median
value of $f_{\sax}/f_{\rosat}$ is 0.4. Fig. \ref{fluxes} includes also the BL
Lacs studied in Paper I. Most sources have $f_{\sax} < f_{ \rosat}$, as shown
by the fact that the median value for all 12 sources is $f_{\sax}/f_{\rosat}$
is 0.4. Fig. \ref{fluxes} should be compared with Fig. 2 of Wolter et
al. (1998; see also Beckmann et al. 2002), which plots the \sax 1 keV fluxes versus
the corresponding \rosat~fluxes for a sample of 8 HBL. There the two
fluxes are within 30 per cent for most sources and the points follow more
closely the line of equal fluxes. Note that the median value of the flux ratio
at 1 keV is $\sim 0.5$ for the 4 sources studied in this paper (and 
$\sim 0.6$ if one includes those studied in Paper I). 
    
Figure \ref{alpha_sax_rosat} shows the \sax spectral index ($0.1-10$ keV)
vs. the \rosat~spectral index ($0.1-2.4$ keV). The larger \sax error bars for
most of our sources, as compared to \rosat, are due to the worse photon
statistics. (The PSPC count rates, in fact, are typically a factor of 20
larger than the LECS ones.) All sources studied in this paper 
occupy the region of the plot where
$\alpha_{\rm x}(\sax) < \alpha_{\rm x}(\rosat)$. The interpretation of this
plot is complicated by variability effects, which affect the shape of the
X-ray spectrum, and possibly by \rosat~miscalibrations (e.g., Iwasawa, Fabian
\& Nandra 1999). Keeping this in mind, in agreement with our previous results
the figure suggests a concave overall X-ray spectrum for our sources, with a
flatter component emerging at higher energies. We find $\alpha_{\rm x}(\rosat)
- \alpha_{\rm x}(\sax) = 0.7\pm0.1$. This difference can hardly be explained
by miscalibration effects which, if present, should only steepen the \rosat~
slopes by $\sim 0.2 - 0.3$ (see also Mineo et al. 2000). Fig.
\ref{alpha_sax_rosat} includes also the BL Lacs studied in Paper I. Note that
only one BL Lac has $\alpha_{\rm x}(\rosat) < \alpha_{\rm x}(\sax)$ and that
for all 12 sources we find $\alpha_{\rm x}(\rosat) - \alpha_{\rm x}(\sax) =
0.53\pm0.16$ ($0.53\pm0.18$ excluding the two HBL).

Again, this figure should be compared with Fig. 3 of Wolter et al. (1998),
which shows the same plot for a sample of 8 HBL. In that case the \sax and
\rosat~spectral indices agree within the errors for all but one source. 
A sample of 10 HBL studied by Beckmann et al. (2002) shows more scatter but
the mean values for the spectral indices are still similar.

Looking at the differences between \sax and \rosat~spectra in more detail, OQ
530 shows, as discussed above (see Tab. \ref{saxbkn}), evidence of a steeper
\sax spectrum at low energies (very significant for the data presented by
Tagliaferri et al. 2003), consistent with the \rosat~data. Similarly, the
\rosat~spectrum of 3C 371 is best fitted by a broken power-law with a hard
component consistent with the \sax spectrum and a steep, soft component which
dominates only for $E \la 0.3$ keV. Could a \rosat-like component be also
present in the \sax spectra of our other two sources but not have been
detected? The answer is: no. We have simulated this by assuming a broken
power-law model with a break at 2.4 keV (the end of the \rosat~band) and a
hard component with spectral index equal to that measured by {\it BeppoSAX}. 
A soft
component with spectral index equal to that seen by \rosat~can be excluded
with very strong significance ($> 99$ per cent) for both AO 0235+164 and S5
1803+784. If the flux were also constrained to be the same as the \rosat~one
the significance of the exclusion would be even higher. In other words, \sax
did not detect a steep, \rosat-like component in these two sources because 
it was not there.
 
\section{Spectral Energy Distributions} 
To address the relevance of our \sax data in terms of emission processes in BL
Lacs, we have assembled multifrequency data for all our sources. The main
source of information was NED, and so most data are not simultaneous with our
\sax observations. For all our sources, however, we were able to find
nearly-simultaneous (typically within a month) radio observations in the
University of Michigan Radio Astronomy Observatory (UMRAO) database. These are
reported in Table \ref{radioobs}, which also gives the nearly-simultaneous
radio-X-ray spectral index, $\alpha_{\rm rx}$ with its error. This is defined
between the rest-frame frequencies of 4.8 GHz and 1 keV, and has been
K-corrected using the X-ray spectral indices given in Tab. \ref{saxfits} and
radio spectral indices between 4.8 and 8.0 GHz derived from the UMRAO data.
One of our sources (AO 0235+164) has been detected by EGRET so its energy
distribution reaches $\sim 5 \times 10^{24}$ Hz. The EGRET data come from the
compilation of Lin et al. (1999), which include the first entries in the Third
EGRET Catalog. 

\begin{figure*}
\centerline{\psfig{figure=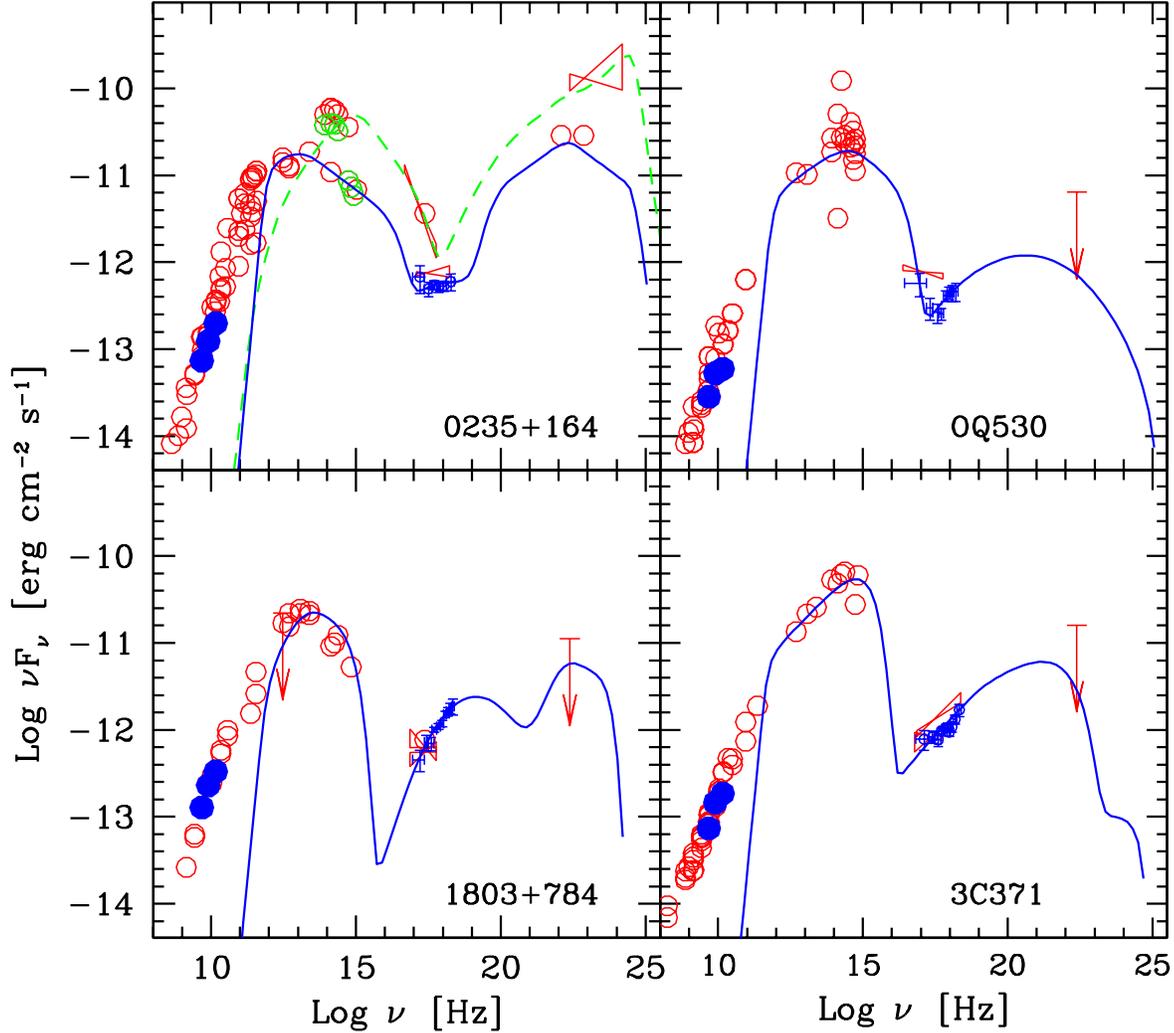,width=17cm}}
\vskip -2.0 true cm
\caption{Spectral energy distributions for our sources. Filled symbols
indicate \sax data and nearly-simultaneous radio data from UMRAO, while open
symbols indicate data from NED. The solid lines correspond to the one--zone
homogeneous synchrotron and inverse Compton model calculated as explained in
the text, with the parameters listed in Table \ref{model}. The dashed line
represents the fit to the high state of AO 0235+164. \rosat, ASCA (for AO
0235+164 and 3C 371), and EGRET (high-state) data are shown by a bow-tie that
represents the spectral index range.\label{seds}}
\end{figure*}

The spectral energy distributions (SEDs) for our sources are shown in
Fig. \ref{seds}, where filled circles indicate \sax data and the
nearly-simultaneous radio data, and open symbols represent non-simultaneous
literature (mostly NED) data. The \sax data have been converted to $\nu f_{\rm
\nu}$ units using the XSPEC unfolded spectra after correcting for absorption.
%
\rosat~(from Tab. \ref{rosatfits} and \ref{rosatbkn}) and EGRET data are shown
by a bow-tie that represents the spectral index range. The plotted \rosat~data
correspond to the fixed galactic column density fits (for AO 0235+164 the one
including the extra absorption), and to the highest and lowest flux observed.
For 3C 371, we plotted the spectral index range from the broken power-law fits
(above 0.3 keV; Tab. \ref{rosatbkn}). We also show the available ASCA data for
AO 0235+164 (Madejski et al. 1996) and 3C 371 (Donato et al. 2001).

To derive the intrinsic physical parameters that could account for the
observed data, we have fitted the SED of our sources with an homogeneous,
one--zone synchrotron inverse Compton model as developed in Ghisellini,
Celotti \& Costamante (2002). This model is very similar to the one described
in detail in Spada et al. (2001; it is the ``one--zone'' version of it), and
is characterized by a finite injection timescale, of the order of the light
crossing time of the emitting region (as occurs, for example, in the internal
shocks scenario, where the dissipation takes place during the collision of two
shells of fluid moving at different speeds).

In this model, the main emission comes from a single zone and a single
population of electrons, with the particle energy distribution determined at
the time $t_{\rm inj}$, i.e., at the end of the injection, which is the time
when the emitted luminosity is maximized. Details of the model can be found in
Ghisellini et al. (2002), who have applied it successfully to both low--power, 
high-peaked BL Lacs and powerful flat-spectrum radio quasars. A summary of the model main
characteristics and a discussion of its application to the SEDs of other \sax
sources can also be found in Padovani et al. (2001; 2002).

Some of our sources have (weak) broad lines and therefore the contribution of
the disc to the SED might not be completely negligible. Furthermore, photons
produced in the broad line region could contribute to the seed photon
distribution for the inverse Compton scattering. We accounted for this by
assuming that a fraction $\sim$ 10\% of the disc luminosity $L_{\rm disc}$ is
reprocessed into line emission by the broad line region (BLR), $L_{\rm BLR}$,
assumed to be located at $R_{\rm BLR}$. $L_{\rm BLR}$ estimates for three of
our sources were found in Celotti et al. (1997). $R_{\rm BLR}$ was assumed to
scale as $L_{\rm disc}^{0.7}$, following Kaspi et al. (2000) and disc 
emission was assumed to be a simple black--body peaking at $10^{15}$ Hz.

The source is assumed to emit an intrinsic luminosity $L^\prime$ and to be
observed with the viewing angle $\theta$. The input parameters are listed in
Table \ref{model}, which gives the name of the source in column 1, $L^\prime$
in column 2, $L_{\rm disc}$ in column 3, $R_{\rm BLR}$ in column 4, the
magnetic field $B$ in column 5, the size of the region $R$ in column 6, the
Lorentz factor $\Gamma$ in column 7, the angle $\theta$ in column 8, the slope
of the particle distribution $n$ in column 9, the minimum Lorentz factor of
the injected electrons $\gamma_{\rm min}$ in column 10, the Lorentz factor of
the electrons emitting most of the radiation $\gamma_{\rm peak}$ in column 11,
and finally $\nu_{\rm peak}^{\rm syn}$ in column 12. Note that $\gamma_{\rm
peak}$ and $\nu_{\rm peak}^{\rm syn}$ are derived quantities and not input
parameters.

The model fits are shown in Fig. \ref{seds} as solid lines. The applied model
is aimed at reproducing the spectrum originating in a limited part of the jet,
thought to be responsible for most of the emission. This region is necessarily
compact, since it must account for the fast variability shown by all blazars,
especially at high frequencies. The radio emission from this compact regions
is strongly self--absorbed, and the model cannot account for the observed
radio flux. This explains why the radio data are systematically above the
model fits in the figures. For AO 0235+164 we have also modelled a high state
corresponding to the hardest EGRET spectrum, which can be accounted for by
assuming also a high X-ray state ({\it ROSAT} data). This is not compatible
(in our model) with the \sax data, which instead correspond to the lowest
X-ray state for this source, and can be fitted well together with the
low-state EGRET data, taken quasi-simultaneously with the ASCA observation
(whose flux and spectrum are very similar to the \sax ones; Madejski et
al. 1996).

As shown in Table \ref{model}, the source dimensions, magnetic field, bulk
Lorentz factors, and viewing angles are quite similar for all sources. The
need for external seed photons for some sources, while indicative of a broad
line region, is not extremely compelling, since the disc luminosities are much
smaller than those required in radio--loud quasars (see, e.g., Ghisellini et
al. 1998). The main difference between sources are in the derived value of
$\gamma_{\rm peak}$ and intrinsic luminosities.

Fig. \ref{seds} shows that the \sax band is dominated by inverse Compton
emission for S5 1803+784 and 3C 371, while it is in between the synchrotron
and inverse Compton regions for AO 0235+164 and OQ 530. These results are
consistent with the \sax data, as discussed in Sect. \ref{fitssax}.

\begin{table*}
\caption{Nearly-simultaneous Radio Observations.\label{radioobs}}
\begin{tabular}{lrlrlrlc}
\hline
Name &F$_{\rm 4.8GHz}$&Observing date&F$_{\rm 8.0GHz}$&Observing date&F$_{\rm 14
.5GHz}$&Observing date&$\alpha_{\rm rx}$ \\
     &              (Jy)&              &              (Jy)&              &
 (Jy)& & \\
\hline
AO 0235+164&$1.54\pm0.03$&1999 Jan 30&$1.55\pm0.08$&1999 Jan 26&$1.37\pm0.03$
&1999 Jan 22&$0.86\pm0.02$\\
OQ 530     &$0.59\pm0.08$&1998 Sep 24&$0.66\pm0.06$&1999 Feb 15&$0.41\pm0.01$
&1999 Mar 17&$0.88\pm0.02$\\
S5 1803+784&$2.67\pm0.03$&1998 Sep 25&$2.88\pm0.17$&1998 Sep 18&$2.29\pm0.20$
&1998 Sep 30&$0.90\pm0.01$\\
3C 371     &$1.53\pm0.03$&1998 Sep 26&$1.82\pm0.18$&1998 Sep 18&$1.28\pm0.03$ 
&1998 Sep 9 &$0.87\pm0.01$ \\
\hline
\end{tabular}
\end{table*}

\begin{table*}
\caption{Model Parameters.\label{model}}
\begin{center}
\begin{tabular}{llllllllllll}
\hline
Name  &$L^\prime$       &$L_{disc}$   &$R_{BLR}$ &$B$ &$R$    &$\Gamma$ &$\theta$ &$n$  &$\gamma_{min}$ &$\gamma_{peak}$ &
 $\nu^{syn}_{peak}$ \\
      & erg s$^{-1}$    &erg s$^{-1}$ & cm       &G   & cm    &         &         &     &          &      &      Hz     \\
\hline
AO 0235$+$164  &1.1e43  &1.3e45      &4.2e17    &3.8  &4.1e16    &16       &2.9     &3.5  &20    &276  &  2.1e13   \\
(high state)  &6.3e43  &1.3e45      &4.2e17    &1.7  &2.3e16    &13       &2.6     &3.9  &3600   & 3600  & 1.5e15   \\
OQ 530         &8.0e41  &...         &...        &3.8  &1.25e16  &12       &4.9     &3.5  &70   &1410  & 3.3e14   \\
S5 1803+784    &6.8e42  &3.9e45      &8.8e17     &5.0  &2.0e16   &11       &3.5     &3.5  &230  &373   &  3.9e13  \\   
3C 371         &2.2e41  &4.8e42      &1.1e16     &1.8  &9.0e15   &13       &5.0     &3.3  &10   &5000  & 1.9e15   \\
\hline 
\end{tabular}
\end{center}
\end{table*}

\begin{figure}
\centerline{\psfig{figure=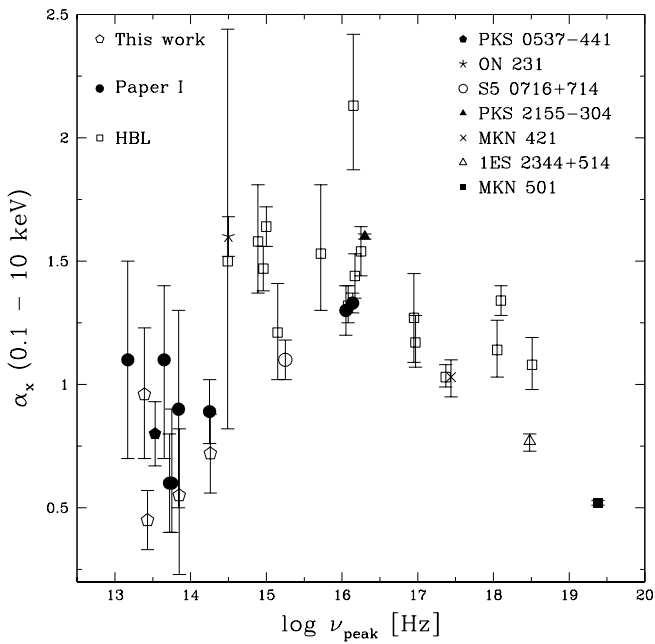,width=9cm}}
\caption{The \sax~spectral index ($0.1-10$ keV) vs. the logarithm of the peak
frequency for our sources (open pentagons), those studied in Paper I (filled
circles), the HBL studied by Wolter et al. (1998) and Beckmann et al. (2002)
(open squares), PKS 0537$-$441 (filled pentagon), ON 231 (star), S5 0716+714
(open circle), PKS 2155$-$304 (filled triangle), MKN 421 (cross), 1ES 2344+514
(open triangle), and MKN 501 (filled square). See text for details and
references.\label{nupeak_alpha}}
\end{figure}

\subsection{X-ray Spectral Index and the Synchrotron Peak Frequency}

One of the aims of this project was to study the dependence of the X-ray
spectral index on the synchrotron peak frequency $\nu_{\rm peak}$ found by
Padovani \& Giommi (1996) and Lamer et al. (1996) from \rosat~data by using the
broader \sax energy band. Padovani \& Giommi (1996) found a strong
anti-correlation between $\alpha_{\rm x}$ and $\nu_{\rm peak}$ for HBL (i.e.,
the higher the peak frequency, the flatter the spectrum), while basically no
correlation was found for LBL. This was interpreted as due to the tail of the
synchrotron component becoming increasingly dominant in the \rosat~band as
$\nu_{\rm peak}$ moves closer to the X-ray band (see Fig. 7 of Padovani \&
Giommi 1996). 

The updated \sax version of this dependence is shown in
Fig. \ref{nupeak_alpha}, which plots the \sax~spectral index ($0.1-10$ keV)
vs. the logarithm of the peak frequency for our sources (open pentagons), the
other 1-Jy sources studied in Paper I (filled circles), the HBL studied by
Wolter et al. (1998) and Beckmann et al. (2002) (open squares), and other BL
Lacs studied by {\it BeppoSAX}. These include, in order of increasing peak
frequency: PKS 0537$-$441 (filled pentagon; Pian et al. 2002), ON 231 (star;
$\alpha_{\rm x}$ in the $0.1-3.8$ keV range; Tagliaferri et al. 2000), S5
0716+714 (open circle; Giommi et al. 1999), PKS 2155$-$304 (filled triangle;
Giommi et al. 1998), MKN 421 (cross; $\alpha_{\rm x}$ in the $0.1-1.6$ keV;
Fossati et al. 2000), 1ES 2344+514 (open triangle; Giommi, Padovani \& Perlman
2000), and MKN 501 (filled square; Pian et al. 1998). When more than a value
of $\alpha_{\rm x}$ was available for these BL Lacs we picked the one
corresponding to the largest $\nu_{\rm peak}$. The $\nu_{\rm peak}$ values for
the sources studied in this paper have been taken from Sambruna, Maraschi \&
Urry (1996), who fitted a parabola to the $\nu f_{\rm \nu}$ broad-band
spectra. The $\nu_{\rm peak}$ values for the HBL studied by Wolter et
al. (1998) and Beckmann et al. (2002) are taken from those papers and
similarly the values for the additional sources are taken from the
referenced papers.

Fig. \ref{nupeak_alpha}, although with less statistics, basically confirms the
\rosat~findings, namely a strong anti-correlation between $\alpha_{\rm x}$ and
$\nu_{\rm peak}$ for HBL and no correlation for LBL. A few differences,
however, are worth mentioning. First, the range in $\alpha_{\rm x}$ is
somewhat smaller ($\sim 1.5$ vs. $\sim 3$). This is likely due to the larger
energy range over which $\alpha_{\rm x}$ is measured ($0.1-10$ keV for \sax
vs. $0.1-2.4$ keV for \rosat). Objects with very steep \rosat~$\alpha_{\rm
x}$, in fact, are those in which synchrotron emission is nearing the
exponential cut-off; by having a larger band \sax includes flatter, higher
energy emission due to inverse Compton. Second, the wide $0.1-100$ keV
coverage of \sax has allowed the detection of spectacular spectral variability
with $\nu_{\rm peak}$ reaching $\ga 10$ keV. As predicted by Padovani \&
Giommi (1996), these sources display very flat $\alpha_{\rm x}$ ($\sim 0.5 -
0.8$), since \sax is sampling the top of the synchrotron emission. Note that
in this case the flat X-ray spectrum is {\it not} associated with inverse
Compton emission, although extreme HBL (objects to the far right in
Fig. \ref{nupeak_alpha}) have X-ray spectra as flat as extreme LBL (objects to
the far left of the figure).

\section{The 1-Jy BL Lac sample: the \sax view}

\sax data are now available for fourteen 1-Jy BL Lacs: four studied here,
seven studied in Paper I, plus MKN 501 (Pian et al. 1998), S5 0716+714 (Giommi
et al. 1999), and PKS 0537$-$441 (Pian et al. 2002). Based on their
$\alpha_{\rm rx}$ and $\nu_{\rm peak}$ values, 11 objects can be classified as
LBL and 3 as HBL, including S5 0716+714 which is an HBL/intermediate
source. We note that these sources include {\it all} 1-Jy BL Lacs with $0.1 -
10$ keV X-ray flux larger than $3 \times 10^{-12}$ erg cm$^{-2}$ s$^{-1}$
(estimated from an extrapolation of the single power-law fits derived for
these objects from \rosat~data; Urry et al. 1996). This sub-sample is
therefore well-defined, makes up a sizeable fraction ($\sim 41$ per cent) of
the full sample, and can therefore be used to obtain robust results on the
hard X-ray emission of X-ray bright 1-Jy BL Lacs.  Radio data from the UMRAO
database are available for ten out of fourteen sources, including MKN 501 and
S5 0716+714. Therefore, $\sim 71$ per cent of our BL Lacs ($\sim 73$ per cent
of LBL) have nearly simultaneous high-frequency radio data which allow us to
study the relationship between X-ray and radio emission almost independently
of variability.

\subsection{The hard X-ray spectra of 1-Jy BL Lacs}

The mean value of the LECS/MECS spectral indices, covering the $0.1 - 10$ keV
range, for the fourteen 1-Jy BL Lacs studied by \sax is $\langle \alpha_{\rm
x} \rangle = 0.83\pm 0.07$. As already indicated by Fig. \ref{nupeak_alpha}, 
however, this value reflects two broadly
different types of sources and X-ray emission processes. 

On one side, in fact, we have the eleven LBL, namely PKS 0048$-$097, AO
0235+164, PKS 0537$-$441, OJ 287, PKS 1144$-$379, OQ 530, PKS 1519$-$273, S5
1803+784, 3C 371, 4C 56.27, and BL Lac. For these sources the X-ray spectrum
is relatively flat, with $\alpha_{\rm x} \la 1$ within the errors and $\langle
\alpha_{\rm x} \rangle = 0.79\pm 0.07$. Five of these sources show 
evidence for a low-energy steepening. While the single $F$--test probabilities
that the improvement provided by a double power-law model is significant range
between 87 and 93 per cent, by adding up the $\chi^2$ values we derive a 98
per cent probability for the five sources together. The \sax picture for LBL,
supported also by their SED, is then that of an inverse Compton dominance,
with the steep synchrotron tail coming in around $\sim 1-2$ keV in about half
the sources. \rosat~data, modulo variability and calibration effects,
corroborate the evidence for overall concave spectra. We find, in fact,
$\langle \alpha_{\rm x}(\rosat) - \alpha_{\rm x}(\sax) \rangle = 0.55\pm0.17$
(and a median value of 0.73). The mean flux ratio ($0.1-2.4$ keV) is $\langle
f_{\sax}/f_{\rosat} \rangle \sim 0.4$ ($\sim 0.3$ median), which is again 
consistent with the picture of a variable synchrotron tail contributing at 
low energy, thus causing both steeper spectra and higher fluxes when
present. This is confirmed also by considering only the five LBL with
indication of low-energy steepening: using the broken power-law fit
values, $\langle f_{\sax}/f_{\rosat} \rangle$ approaches one ($\sim
0.75$, with three objects having values $\sim1$ and two $\sim0.4-0.5$). 

On the other side, we have the three HBL, namely S5 0716+714, MKN 501, and PKS
2005$-$489. The mean value in this case is $\langle \alpha_{\rm x} \rangle =
1.0\pm 0.2$, which increases to $\langle \alpha_{\rm x} \rangle = 1.2\pm 0.1$
if we exclude MKN 501, found in an extremely high state, with very
large $\nu_{\rm peak}$ and a flat synchrotron spectrum, as predicted by the 
correlation shown in Fig. \ref{nupeak_alpha}. The evidence 
for spectral
curvature here is mixed. While MKN 501 shows flattening at lower energies
($\Delta \alpha_{\rm x} \sim -0.2$ below $\sim 2$ keV; Pian et al. 1998), S5
0716+714 shows steepening ($\Delta \alpha_{\rm x} \sim 0.7$ below $\sim 2-3$
keV; Giommi et al. 1999). This difference can be explained by looking at
Fig. \ref{nupeak_alpha}. While S5 0716+714 is an HBL with $\nu_{\rm peak} \sim
10^{15}$ Hz, and therefore relatively close to LBL values, MKN 501 was
observed by \sax in outburst, with $\nu_{\rm peak} \ga 100$ keV or $\ga 2
\times 10^{19}$ Hz. In one case we are then seeing a mix of synchrotron and
inverse Compton radiation, while in the other we are looking at the overall
steepening of pure synchrotron emission. The \sax picture for HBL, supported
also by their SED, is then that of a synchrotron dominance, with the flat
inverse Compton component making a contribution for intermediate sources. 
As regards \rosat~data, we get $\langle \alpha_{\rm x}(\rosat) -
\alpha_{\rm x}(\sax) \rangle = 0.89\pm0.03$ (and a median value of
0.92). The mean flux ratio is $\langle f_{\sax}/f_{\rosat} \rangle \sim
1.5$ ($\sim 1.5$ median), a factor $\sim 4$ larger than for the LBL. 
The mean flux ratio at 1 keV is a factor $\sim 7$ larger for HBL than
for LBL. 


We note that large variability has been seen in the synchrotron component of
several BL Lacs observed with \sax (Ravasio et al. 2002, and references
therein) and possibly only once, to the best of our knowledge, in their
inverse Compton component (in the case of PKS 1144$-$379: Paper I). 
Comparing single power-law fits, however, one should be aware
of another factor that could contribute to different values
of $f_{\sax}/f_{\rosat}$ for HBL and LBL.
As pointed out by Beckmann et al. (2002),
the different X-ray spectral shapes of HBL and LBL could play a role,
given the different spectral response of the two instrument.
LBL have flat inverse Compton spectra but synchrotron emission, with a steep
spectrum, sometimes dominates at soft energies (in $\sim 45$ per cent of our
sources). The fits to their \sax LECS/MECS spectra, which sample a much
larger range towards high energies than the \rosat~PSPC spectra, may then be
mostly sensitive to the flat inverse Compton component, with the result that
the fitted flux at 1 keV (and integrated in the soft band) 
is lower than that inferred from the \rosat~fits. This effect is less
important when a broken power-law is adopted (see above), but the relatively
low statistics of the LECS data  
may not always justify a statistical preference of this model
with respect to the single power-law one.


\subsection{X-ray and radio powers and $\alpha_{\rm rx}$}

\begin{figure}
\centerline{\psfig{figure=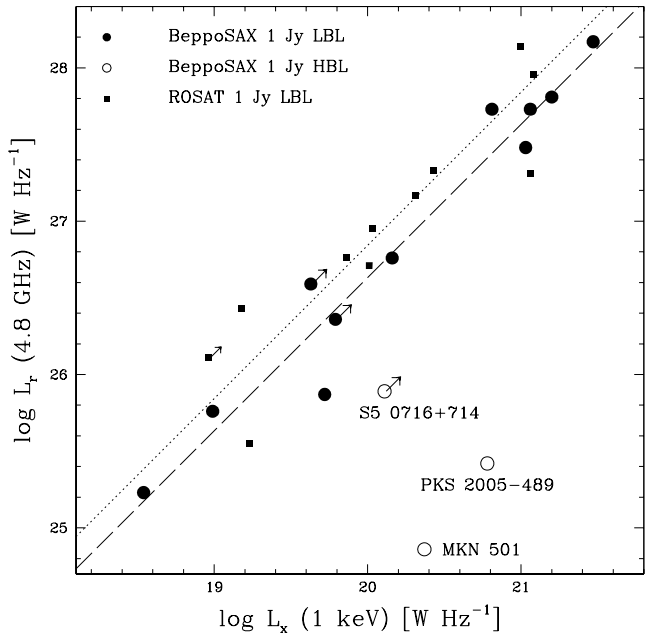,width=9cm}}
\caption{Nearly simultaneous X-ray and radio luminosities of 1-Jy BL Lacs
objects observed by {\it BeppoSAX} (circles) and {\it ROSAT} (squares). Lower
limits indicate sources for which no redshift is available but $z > 0.2$ was
assumed, based on the fact that the host galaxies are optically unresolved.
The dashed line indicates a value of $\alpha_{\rm rx} = 0.861$, the mean value
for the \sax LBL, while the dotted line indicates a value of $\alpha_{\rm rx}
= 0.89$, the mean value for the {\it ROSAT} LBL. Labels identify the
HBL (open circles).\label{lxlr}}
\end{figure}

As mentioned above, since $\sim 70$ per cent of the objects have nearly
simultaneous X-ray and radio data, we can address the relation between X-ray
and radio emission almost independently of variability. Fig. \ref{lxlr} plots
the radio power $L_{\rm r}$ vs. the X-ray power $L_{\rm x}$ for our sources.
As expected, the HBL, having larger $\nu_{\rm peak}$ values, have larger
$L_{\rm x}$ at a given $L_{\rm r}$ and therefore occupy the bottom-right part
of the diagram. The correlation between $L_{\rm r}$ and $L_{\rm x}$ for LBL is
surprisingly tight, strong, and linear. We find $L_{\rm r} \propto L_{\rm
x}^{0.99 \pm 0.08}$, significant at the $> 99.99$ per cent level. A partial
correlation analysis (e.g., Padovani 1992), shows that the correlation is
still very strong ($P \sim 99.7$ per cent) even when the common redshift
dependence of the two powers is subtracted off. Equivalently, $\alpha_{\rm
rx}$ has a very small dispersion for LBL. We get $\langle \alpha_{\rm rx}
\rangle = 0.861 \pm 0.008$ (dashed line in Fig. \ref{lxlr}). Given the
tightness of the correlation, and the importance that the suggested strong
link between the radio and X-ray band would have for LBL, we have looked in
the UMRAO database for radio observations at 4.8 GHz nearly simultaneous with
the {\it ROSAT} data presented by Urry et al. (1996). This provided us with 11
more sources, for a total of 22 1-Jy LBL, $\sim 86$ per cent of them with
nearly simultaneous X-ray and radio data. The results obtained for our \sax
sources are confirmed with this enlarged sample. Namely, we find now $L_{\rm
r} \propto L_{\rm x}^{0.97 \pm 0.08}$, significant at the $> 99.99$ per cent
level. As before, the correlation is still very strong ($P \sim 99.97$ per
cent) when the common redshift dependence is subtracted off. We get $\langle
\alpha_{\rm rx} \rangle = 0.89 \pm 0.01$ (dotted line in Fig. \ref{lxlr}) for
the 11 {\it ROSAT} sources, not significantly different from our previous
value, and a mean for the 22 1-Jy LBL of $\langle \alpha_{\rm rx} \rangle =
0.875 \pm 0.008$. This means that the X-ray power (or flux) can be
extrapolated from the radio power (or flux) within $\sim 15$ per cent over
almost eight orders of magnitude in frequency. 

It is interesting to note that the strong $L_{\rm r} - L_{\rm x}$ correlation
displayed by the 1-Jy LBL is {\it not} only due to the nearly simultaneous
radio and X-ray data. Even if we use the original 1-Jy data, in fact, we find
a strong correlation ($P > 99.99$ per cent) for the 22 sources in Fig. 6,
still very significant ($P \sim 99.97$ per cent) when the redshift effect is
subtracted off. The slope of the correlation is slightly flatter and $\langle
\alpha_{\rm rx} \rangle$ is slightly larger (although not significantly so:
$0.890\pm0.008$), as on average the 1 Jy fluxes are larger than the UMRAO
ones, but the dispersion is still very small. We note that the relatively
small range in $\langle \alpha_{\rm rx} \rangle$ for LBL, although never
studied with nearly simultaneous radio and X-ray data, has been previously
noted, amongst others, by Padovani \& Giommi (1995) and Fossati et al. (1998).
To the best of our knowledge, however, all previous studies of the $L_{\rm r}
- L_{\rm x}$ correlation for 1-Jy BL Lacs did not distinguish between LBL and
HBL, thereby reducing the significance of the correlation.

The interpretation of the strong linear link between X-ray and radio
emission components in LBL suggested by the $L_{\rm r} - L_{\rm x}$
correlation is not straightforward within the currently accepted
scenario. As illustrated in Fig. \ref{seds}, in fact, our one-zone model
cannot reproduce the observed radio flux, thought to originate in a region
of the jet larger than the (more compact) one responsible for most of the
emission, including X-rays. On the other hand, since the radio spectra in
these sources are invariably flat, there is a correlation between the radio
flux at the self-absorption frequency ($\sim 10^2-10^3$ GHz) of the smaller
X-ray emitting region and the flux at 5 GHz. This may account, at least in
part, for the $L_{\rm r} - L_{\rm x}$ correlation, even if the powers in
the two bands are produced in different parts of the jet.

\section{Conclusions}

We have presented new \sax observations of four BL Lacertae objects selected
from the 1-Jy sample, all LBL, i.e., characterized by a peak in their
multifrequency spectra at infrared/optical energies. We have then used these
data plus data previously published by us (7 sources; Paper I) and others (3
sources) to study the properties of the hard X-ray spectra of an X-ray
flux-limited 1-Jy sub-sample (containing 11 LBL and 3 HBL) and constrain the
emission processes.

A relatively simple picture comes out from this paper: a dominance of inverse
Compton emission in the X-ray band of LBL, with $\sim 50$ per cent of the
sources showing also a likely synchrotron component. Our main results are
as follows:

\begin{enumerate}

\item The \sax spectra of our LBL sources are relatively flat ($\alpha_{\rm x}
\sim 0.8$). For five sources broken power-law models improve the fits at the
98 per cent level when the $\chi^2$ values for the single sources are added
up. The resulting best-fit spectra all concur in indicating a flatter
component emerging at higher energies, with spectral changes $\Delta
\alpha_{\rm x} \sim 0.8$ around $1-2$ keV.

\item Our LBL have a typical difference between \sax and \rosat~spectral
slopes $\alpha_{\rm x}(\rosat) - \alpha_{\rm x}(\sax) \sim 0.6$. The
interpretation of this effect is complicated by possible 
\rosat~miscalibrations (which, if at all present, could explain a 
difference $\sim 0.3$) and variability effects, as the mean flux ratio is 
$\langle f_{\sax}/f_{\rosat} \rangle \sim 0.4$, but is consistent with the 
picture of a variable synchrotron emission contributing in the soft X-ray 
band, leading to high fluxes and steeper spectra when present.
%

\item Despite the non-simultaneity of the multifrequency data (UMRAO radio
data excluded) it is apparent that the \sax spectra indicate a different
emission component in the SEDs of our LBL sources, separate from that
responsible for the low energy emission. In fact, the extrapolation of the
relatively flat \sax slopes cannot be extended to much lower frequencies since
the predicted optical flux would be orders of magnitude below the observed
value. A sharp steepening towards lower frequencies is then necessary to meet
the much higher optical (synchrotron) flux.

\item Our interpretation of the $\alpha_{\rm x} - \nu_{\rm peak}$ diagram is
the one originally proposed by Padovani \& Giommi (1996) for the
\rosat~data. Namely, $\alpha_{\rm x}$ steepens moving from LBL to HBL as
synchrotron replaces inverse Compton as the main emission mechanism in the
X-ray band. The spectral index then flattens again as the synchrotron peak
moves to higher energies in the X-ray band, eventually converging to the
relatively flat value characteristic of synchrotron emission before the
peak. Again, this fits perfectly with a dominance of inverse Compton emission
in our LBL.

\item We find a very tight proportionality between nearly simultaneous radio
and X-ray powers for our LBL sources and 11 additional LBL with {\it ROSAT}
data, such that X-ray power can be predicted within $\sim 15$ per cent from
the radio power. This points to a strong link between X-ray and radio emission
components in LBL. 

\item The data for the (small number of) HBL are consistent with other studies
based on larger samples of sources and confirm the dominance of synchrotron 
emission in the \sax band. 

\end{enumerate}

\section*{Acknowledgements}
We thank Andrea Comastri, Giovanni Fossati, Franco Mantovani, Laura Maraschi,
Carlo Stanghellini, Gianpiero Ta\-glia\-ferri, and Meg Urry for their
contribution at an early stage of this project. 
LC acknowledges the STScI
Visitor Program. This research has made use of data from the University of
Michigan Radio Astronomy Observatory which is supported by funds from the
University of Michigan and of the NASA/IPAC Extragalactic Database (NED),
which is operated by the Jet Propulsion Laboratory, California Institute of
Technology, under contract with the National Aeronautics and Space
Administration.

\end{document}